\documentstyle[12pt,aaspp4]{article}

\def\icm{\hbox{${\rm cm}^{-1}$}}
\def\beq{\begin{equation}}
\def\eeq{\end{equation}}

\begin{document}

\slugcomment{Submitted to Ap. J. Letters; \quad \tt{astro-ph/9612040} }

\title{
Whole Disk Observations of Jupiter, Saturn and Mars in 
Millimeter--Submillimeter Bands
}

\author{
A.~B.~Goldin\altaffilmark{1},
M.~S.~Kowitt\altaffilmark{1},
E.~S.~Cheng\altaffilmark{2},
D.~A.~Cottingham\altaffilmark{3},
D.~J.~Fixsen\altaffilmark{4},
C.~A.~Inman\altaffilmark{2},
S.~S.~Meyer\altaffilmark{1},
J.~L.~Puchalla\altaffilmark{2},
J.~E.~Ruhl\altaffilmark{5},
and~R.~F.~Silverberg\altaffilmark{2}
}
\altaffiltext{1}{Department of Astronomy and Astrophysics, and 
   Enrico Fermi Institute, University of Chicago, 5640 South Ellis Avenue,
   Chicago, IL 60637}
\altaffiltext{2}{Laboratory for Astronomy and Solar Physics,
   NASA/Goddard Space Flight Center, Code 685, Greenbelt, MD 20771}
\altaffiltext{3}{Global Science and Technology, Inc.,
   Laboratory for Astronomy and Solar Physics, 
   NASA/GSFC Code 685, Greenbelt, MD 20771}
\altaffiltext{4}{Hughes STX, Laboratory for Astronomy and Solar Physics, 
   NASA/GSFC Code 685, Greenbelt, MD 20771}
\altaffiltext{5}{Department of Physics, University of California,
   Santa Barbara, CA  93106}

\begin{abstract}
Whole disk brightness ratios for Jupiter, Saturn, and Mars are reported at
5.7, 9.5, 16.4, and 22.5 \icm.  Using models for the brightness temperature
of Mars, the whole disk brightness temperatures for Jupiter and Saturn are
also given for the four frequencies.
\end{abstract}

\section{Introduction}
\label{sec:intro}
Whole-disk brightness temperature measurements of the planets are frequently
used as calibrators for radio and infrared astronomy.  
For instruments with beam size larger than $\sim 1\arcmin$, 
planets are bright, unresolved sources ideal for  mapping the shape of the 
far-field antenna pattern as well as providing an absolute calibration.   
In particular, for studies of anisotropy in the cosmic microwave background 
radiation (CMBR), precise common calibration targets are needed to 
permit comparison of experimental results with each other and with theories.
Jupiter and Saturn, while bright and frequently observable, have complicated 
atmospheres which introduce substantial uncertainties in modeling their 
brightness temperatures.  
Mars, however, has only a tenuous CO$_2$ atmosphere which, for broad-band 
millimeter and submillimeter observations, can be safely neglected compared 
with thermal emission from the Martian surface (\cite{griffin86} and 
\cite{wright76}).

We report here the results of multi-frequency observations of Jupiter, Saturn,
and Mars.  Primary results are whole-disk brightness ratios of Mars to 
Jupiter, and Saturn to Jupiter; using model calculations of the Martian 
brightness temperature, we also present whole disk brightness temperatures
for Jupiter and Saturn.

\section{Instrument and Observation}
\label{sec:obs}
The observations were performed with the Medium Scale Anisotropy Measurement,
MSAM1, a balloon-borne instrument designed to measure anisotropy in the 
CMBR at 0\fdg5 scales.  
This instrument has been described in \cite{fixsen96a};
only the salient features are reviewed here.

MSAM1 uses a 1.4~m off-axis Cassigrain telescope with a nutating secondary 
mirror for beamswitching.  The main lobe of the telescope is 28\arcmin\ FWHM, 
and the beam follows a 2~Hz four phase square wave chop 
(i.e., center, left, center, right) with $\pm$ 40\arcmin\ amplitude.  
The telescope has a multimode feed, instrumented with a 4 band bolometric 
radiometer.  The effective frequency and bandwidths of the
four bands, using Rayleigh-Jeans  color corrections, are given in 
Table~\ref{tab:bands}.  A gyroscope is aligned with the telescope main
beam to provide an inertial reference, while absolute pointing
is determined using a CCD star camera.

\begin{deluxetable}{crr}
\tablecolumns{3}
\tablecaption{Bandpass Integrals over Rayleigh-Jeans Spectrum \tablenotemark{a}
   \label{tab:bands}}
\tablehead{
   \colhead{Band} & $<\nu>$ (\icm ) & $\Delta \nu$ (\icm ) \\
}
\startdata
  1  &  5.73    & 1.32 \nl
  2  &  9.54    & 2.39 \nl
  3  & 16.4\phn & 1.82 \nl
  4  & 22.5\phn & 1.32 \nl
\enddata
\tablenotetext{a}
{
The effective frequency and bandwidth are given by
$<\nu> = \frac{\int \nu F(\nu) I_\nu d\nu}{\int F(\nu) I_\nu d\nu}$,
$\Delta \nu = \frac{\int F(\nu) I_\nu d\nu}{I_{<\nu>}}$, 
where $I_\nu \propto \nu^2$ is a Rayleigh-Jeans spectrum and
$F(\nu)$ is the filter function of the band (see, e.g., \cite{page94}).
}
\end{deluxetable}

The observations were made in a flight from the National Scientific 
Balloon Facility in Palestine, Texas on 1995 June 2.
At the beginning of the flight, we performed a raster
scan across Jupiter to determine the antenna pattern, and then
performed horizontal calibration scans across Jupiter and Mars
for flux calibration.  For the next 5 hours, the telescope
executed deep scans above the North Celestial Pole to measure
CMBR.  After the CMBR observations, the telescope was again
slewed to Jupiter.  Another raster was performed to re-check the
antenna pattern, followed by another calibration scan on
Jupiter.  Finally, a calibration scan was performed on Saturn.
The detailed circumstances for each of the four calibration
scans are given in Table~\ref{tab:circumstances}.  
Because the observations are made from balloon altitudes, atmospheric
extinction is completely negligible.
Of particular note is that, for these observations, the Saturn ring
inclination angle to Earth is nearly zero ($-0\fdg37$); thus, we are
measuring only the disk.

\begin{deluxetable}{rccccc}
\tablecolumns{6}
\tablecaption{Observation Circumstances for 1995 June 2\label{tab:circumstances}}
\tablehead{
  \colhead{Scan} & \colhead{UT} & \colhead{Lat.} & \colhead{Long.}
                 & \colhead{Alt. (km)} & \colhead{Elevation} \\
}
\startdata
 Jupiter--1 & 03:41:51 -- 03:49:12   & 
             31\arcdeg 19\arcmin\ N & 
             95\arcdeg 41\arcmin\ W &
             34.9       & 25\arcdeg 32\arcmin\ -- 26\arcdeg 30\arcmin \nl
  Mars     & 03:59:01 -- 04:11:47   &
             31\arcdeg 17\arcmin\ N &
             95\arcdeg 44\arcmin\ W &
             36.1       & 32\arcdeg 55\arcmin\ -- 30\arcdeg 09\arcmin \nl
 Jupiter--2 & 09:47:05 -- 09:53:49   &
             31\arcdeg 28\arcmin\ N &
             98\arcdeg 27\arcmin\ W &
             37.6       & 18\arcdeg 56\arcmin\ -- 17\arcdeg 48\arcmin \nl
  Saturn   & 10:11:03 -- 10:17:13   &
             31\arcdeg 23\arcmin\ N &
             98\arcdeg 39\arcmin\ W &
             37.3       & 30\arcdeg 07\arcmin\ -- 31\arcdeg 15\arcmin \nl
\enddata
\end{deluxetable}

\section{Data Analysis}
\label{sec:data}

These planet observations have a high signal-to-noise ratio, even in the
raw time-ordered data.  
This requires a somewhat different approach for the data reduction than 
that previously reported for CMBR anisotropy (\cite{cheng94}, \cite{cheng95}, 
and \cite{inman96}). 
Systematic effects, particularly with regards to precise telescope
pointing, are the limiting factors here.

\subsection{Pointing}

The accurate determination of the telescope orientation is critical for
this measurement.
We find the telescope pointing by matching star camera images against a
star catalog.  This fixes the position of the camera frame at the time of
the exposure; between exposures, pointing is interpolated with the gyroscope
outputs.  
Typically, the pointing drifts about 2\arcmin\ between successive camera 
exposures.  The residual errors from interpolating between exposures is 
presumably several times smaller than this.
The position of the main telescope beam within the camera frame is determined
from the raster scan across Jupiter.
Note that for the planetary observations, the image of the planet
itself is deleted from the CCD frame, and background stars are used to
establish the celestial coordinates of the beam.  This ensures that blooming
in the CCD due to the bright planet does not compromise the attitude solution.  
Noise in the gyroscope readout leads to a random RMS pointing uncertainty of 
0\farcm7.

\subsection{Detector Data Reduction}
\label{reduction}

In this analysis, we use a ``double-difference'' demodulation of the data.
For each complete cycle of the secondary mirror, the data for the two side 
beams are averaged and subtracted from the average of the central beam data, 
producing a single demodulated value every 0.5~s for each of the four 
radiometer bands.  This results in a symmetrical, three-lobe antenna 
pattern that is well suited to absolute flux determinations.  

Slow offset drifts are present in the data, and must be removed.  For each
observation listed in Table~\ref{tab:circumstances}, a single linear drift in 
time is fit to those portions of the data corresponding to times when the
telescope was pointed well away from the target planet.  This linear drift
is then subtracted from the data.  Since the observations are short and the
drifts are slow, this simple model is adequate for dedrifting.

The detector signal contains transient spikes due to cosmic rays striking 
the detectors.  
In our previous analyses, the very low instantaneous signal-to-noise 
permitted the identification and removal
of these transients directly from the time-ordered data.  The presence of
large signals from the target planets prevents this procedure; instead, a
smooth spatial model is fit to the data, and cosmic ray spikes are identified
as significant outliers from the fit.  For the raster observations, each of
the 9 horizontal scans was fit to a cubic spline with 30 uniformly-spaced 
knots.  An initial noise estimate $\sigma$ is formed from the RMS of the
residuals from the fit, and then $3\sigma$ outliers are deleted.  The fit
is then repeated, and vertical splines are used to interpolate between the 
scan lines to form a $2\fdg8 \times 0\fdg9$ beammap.  Finally, raw detector
noise is estimated from the RMS of a subset of the data pointed at least 
1\fdg4 away from the target planet.

The calibration scans were analyzed in a similar way, except that instead of
free splines, the fit model was constructed from the beammap derived in the
raster analysis.  For each datum in a calibration scan, the telescope 
pointing is used to determine a planetocentric $X,Y$ coordinate, which is then 
referenced to obtain the beammap amplitude.  A single free parameter, the 
overall scan-to-raster flux ratio, is then fit to the data.  
Again, $3\sigma$ outliers are deleted, and the fit is repeated.  
Between 2\% and 8\% of the data are removed this way, depending on scan.
For this procedure, the early Jupiter raster is used for fitting the 
two scans at the beginning of the flight, while the late Jupiter raster 
is used for the two scans at the end of the flight.
A systematic check on this processing is provided by the Jupiter calibration
scans, which should yield a scan-to-raster flux ratio of 1 (within errors).
The fit results are given in Table~\ref{tab:fratio}.

\begin{deluxetable}{cllll}
\tablecolumns{5}
\tablecaption{Ratios of Target Planet Flux to Jupiter Flux 
   \label{tab:fratio}}
\tablehead{
   \colhead{Band} & 
   \colhead{Jupiter--1} & 
   \colhead{Mars} & 
   \colhead{Jupiter--2} & 
   \colhead{Saturn} }
\startdata
  1  &  $1.036\pm 0.016$   & $(256\pm 2.2)\times 10^{-4}$ & $0.979\pm 0.016$ &
  $ (111\pm 1.0)\times 10^{-3}$\nl
  2  &  $1.016\pm 0.015$   & $(263\pm 2.3)\times 10^{-4}$ & $0.994\pm 0.015$ &
  $ (101 \pm 0.9)\times 10^{-3} $\nl
  3  &  $1.011\pm 0.016$   & $(325\pm 2.8)\times 10^{-4}$ & $0.993 \pm 0.016$ &
  $ (109\pm 0.9)\times 10^{-3}  $\nl
  4  &  $0.985\pm 0.019$   & $(335\pm 2.9)\times 10^{-4}$ & $1.044 \pm 0.016$ & 
  $ (112\pm 1.2)\times 10^{-3}   $ \nl
\enddata
\end{deluxetable}

\subsection{Error Analysis}
\label{errors}

Detailed Monte-Carlo simulations were used to estimate the errors on the 
scan-to-raster flux ratios.  Each realization was generated by starting
with the measured bolometer and pointing data, and adding normally 
distributed random numbers with variances corresponding to the estimated
bolometer and position-readout noise, respectively.  An additional random
linear position drift, corresponding to the slow absolute pointing uncertainty,
was also added to the simulated pointing data.  
Note that pointing noise and drift were simulated in both azimuth and 
elevation.  For each simulated dataset,
we reconstruct the beammaps and scan-to-raster flux ratios according to
the procedure described above.  Final error estimates for the scan-to-raster
flux ratios are determined from the standard deviation of the simulated
ratios.  In contrast to our CMBR measurements, here we find that the 
uncertainty is dominated by the position readout noise, and not the bolometer
noise.

We note the presence among the Jupiter scan results in Table~\ref{tab:fratio} 
of two values out of eight (Jupiter--1 band 1, and Jupiter--2 band 4) 
with greater than $2\sigma$ deviations from unity.
This has $\sim 3\%$ probability (based on $\chi^2=17$ for 8 degrees of 
freedom), and so may be evidence of an additional unaccounted systematic 
error in the data.  We have conservatively decided to inflate the estimated 
uncertainties that follow by a factor of 1.46, which forces the 
reduced $\chi^2$ of the Jupiter scans to unity.

\begin{deluxetable}{cll}
\tablecolumns{3}
\tablecaption{Ratios of Target Planets Temperature to Jupiter Temperature
   \label{tab:tratio}}
\tablehead{
   \colhead{Band} &  \colhead{Mars} & \colhead{Saturn} }
\startdata
  1   & $1.158\pm 0.015$  & $0.833\pm 0.012$\nl
  2   & $1.189\pm 0.015$  & $0.758\pm 0.010$\nl
  3   & $1.470\pm 0.019$  & $0.818\pm 0.012$\nl
  4   & $1.515\pm 0.019$  & $0.840\pm 0.012$ \nl
\enddata
\end{deluxetable}

\subsection{Whole-Disk Brightness Temperature Ratios}
The raw flux ratios determined in \S\ref{reduction} are converted to
whole-disk brightness temperature ratios using the effective mm/sub-mm band
planetary equatorial radii and ellipticities of \cite{hildebrand85}: 
$R_{\rm eq}$ $(\varepsilon) =$ 3397 (0.006), 71495 (0.065), and 60233~km (0.096)
for Mars, Jupiter, and Saturn, respectively, along with
their geocentric distances  and polar inclinations at the epoch of observation.

These constitute the primary results reported here, and are given in 
Table~\ref{tab:tratio}.  The errors reported in the table reflect the 
total uncertainty, as determined in \S\ref{errors}, and include the extra
scale factor (1.46).

\section{Mars Models}
\label{sec:models}
To convert the brightness ratios into whole-disk brightness
temperatures, an epoch-dependent thermal model of Mars is needed to provide 
an absolute calibration.  We have used two distinct models for this purpose.

The first model considered (\cite{wright76}, \cite{wright80}) is
extensively used in the literature (see, e.g., \cite{hildebrand85} and
\cite{griffin86}), and is based on 10--20\micron\ radiometer observations 
of Mars by the {\sl Mariner 6 \& 7} spacecraft.  We follow the example
of \cite{hildebrand85}, and truncate this model assuming 
$T(\lambda \ge 350 \micron) = T(\lambda = 350 \micron)$.  The estimated
model uncertainty at long wavelengths is $\pm 10$K (\cite{wright76}).

The second model used (\cite{rudy87a}, \cite{rudy87b}, \cite{muhleman91})
is based on a physical model of the dielectric properties of the upper 
meter of the Martian surface, constrained by polarized flux measurements
obtained from ${\sl VLA}$ observations at $\lambda = 2$ and 6~cm.  Some
extrapolation is needed to estimate the properties of the regolith
at our wavelengths; we have assumed a dielectric constant 
$\epsilon = 2.25 \pm 0.25$, and a power absorption length 
$l_\nu = (11 \pm 4)\lambda$ (\cite{muhleman91}).  The model uncertainty due
to the input parameter uncertainties ($\pm 3$~K) is somewhat smaller than 
that due to neglecting the effect of scattering (estimated at $\lambda=2.7$~mm
to be $\pm 6$~K), giving a total model uncertainty of $\pm 7$~K.  

Both of these models involve substantial extrapolations in 
wavelength to reach our bands.  We find it reassuring, however, that the two 
models, tuned to substantially different observations at very different
wavelengths, give predicted brightness temperatures that agree well 
within their estimated uncertainties.  

\section{Brightness Temperatures}
The modeled Mars temperatures, along with the derived whole-disk brightness 
temperatures for Jupiter and Saturn, are presented for both models in 
Table~\ref{tab:temperatures}.
Note that the errors listed reflect the uncertainty in the brightness ratios
described in \S\ref{errors}, but do not include the $\sim$5\% Mars model 
uncertainties which are common to all the points.

\begin{deluxetable}{ccccccc}
\tablecolumns{7}
\tablecaption{Temperatures of planets. 
   \label{tab:temperatures}}
\tablehead{
   \colhead{} & 
\multicolumn{3}{c}{ Wright model ( $T=T_{350\mu m} $)} &
\multicolumn{3}{c}{ Rudy model (extrapolated)}\\
\cline{2-4} \cline{5-7}\\
\colhead{Band} 
& \colhead{Mars} &  \colhead{Jupiter} & \colhead{Saturn}
& \colhead{Mars} &  \colhead{Jupiter} & \colhead{Saturn}
 }
\startdata 
  1 & 196 & $169\pm 2$ & $141\pm 3$ & 196 & $169\pm 2$ & $141\pm 3$\nl
  2 & 196 & $165\pm 2$ & $125\pm 2$ & 198 & $166\pm 2$ & $126\pm 2$\nl
  3 & 196 & $133\pm 2$ & $109\pm 2$ & 201 & $137\pm 2$ & $112\pm 2$\nl
  4 & 196 & $129\pm 2$ & $109\pm 2$ & 203 & $134\pm 2$ & $112\pm 2$\nl
\enddata
\tablenotetext{} {The tabulated errors do not include the Mars model 
uncertainty of $\pm 10$K (Wright) or $\pm 7$K (Rudy).}
\end{deluxetable}

\begin{figure}[tbhp]
\plotone{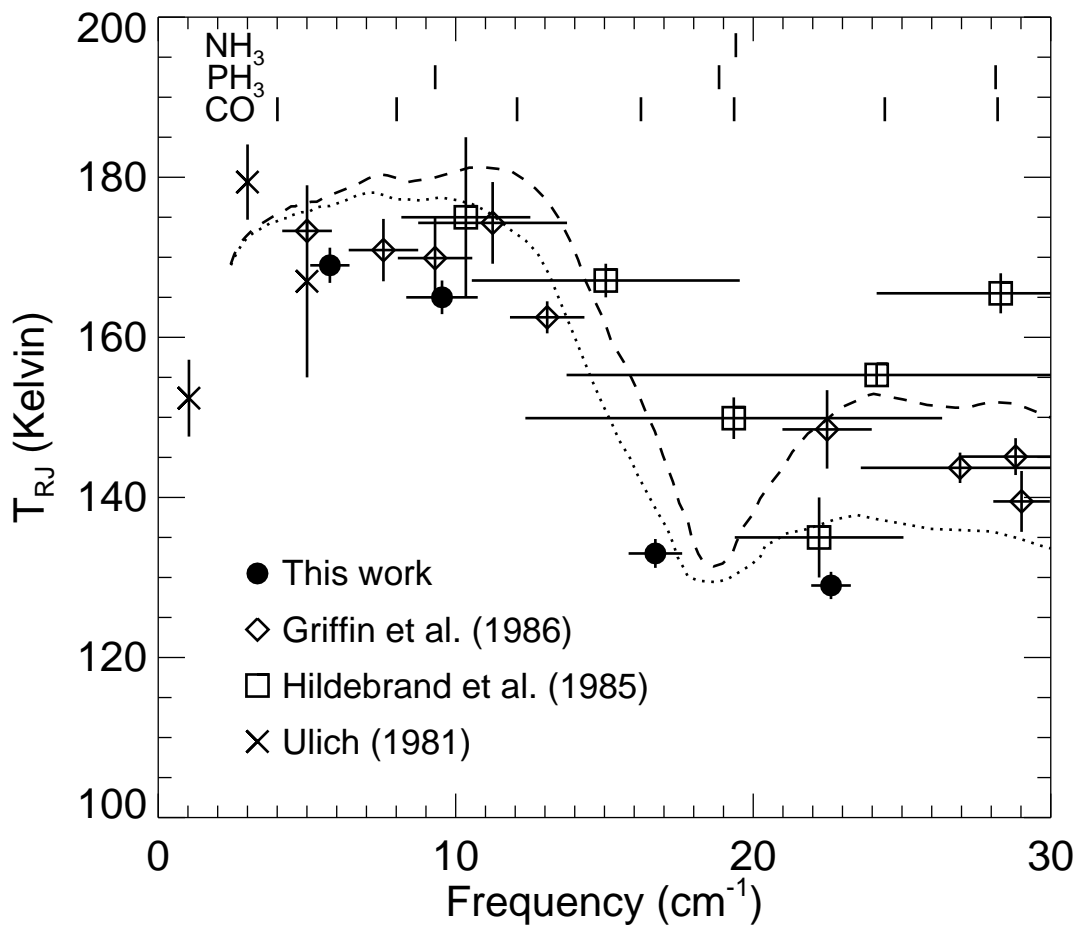} \\[3ex]
\caption[ ]{ Whole-disk brightness temperatures for Jupiter. 
   The Wright thermal model of Mars is used to determine the calibration.
   The vertical bars reflect the relative photometry errors, while the 
   horizontal bars show the bandwidth of the measurements.
   The plotted errors do not include the Mars model uncertainty of $\pm 5\%$.
   The model spectra shown are from \protect\cite{griffin86}, and assume
   clear-sky (dashed) or NH$_3$ cloud with particle size 100~$\mu$m and a 
   particle scale height to gas scale height ratio of 0.15 (dotted).
   The molecular lines from which they are calculated are shown at the top of
   the figure.  \label{fig:jupiter} }
\end{figure}

\begin{figure}[tbhp]
\plotone{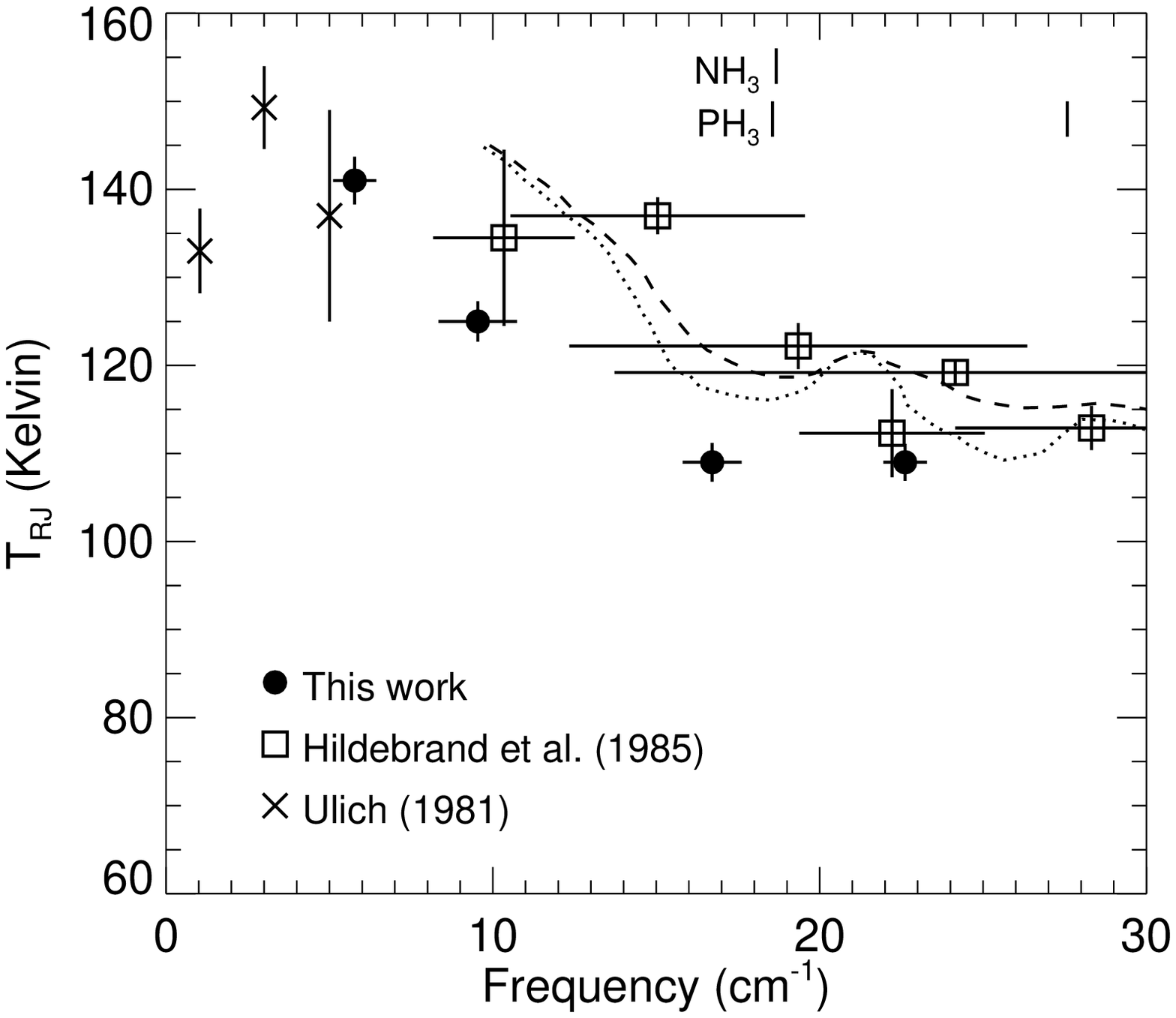} \\[3ex]
\caption[ ]{Whole-disk brightness temperatures for Saturn. 
   The Wright thermal model of Mars is used to determine the calibration.
   The vertical bars reflect the relative photometry errors, while the 
   horizontal bars show the bandwidth of the measurements.
   The plotted errors do not include the Mars model uncertainty of $\pm 5\%$.
   The model spectra shown are from \protect\cite{hildebrand85}, and assume 
   an NH$_3$ mixing ratio of $2\times 10^{-4}$ in the deep atmosphere, and
   PH$_3$ mixing ratios equal to $1.5\times 10^{-6}$ (dashed) or 
   $1.0\times 10^{-5}$ (dotted).
   The molecular lines from which they are calculated are shown at the top of
   the figure.  \label{fig:saturn} }
\end{figure}

Figures \ref{fig:jupiter} and \ref{fig:saturn} plot the brightness temperatures
for Jupiter and Saturn using the Wright model for Mars; this is chosen to
permit easy comparison with the earlier measurements of \cite{ulich81},
\cite{hildebrand85}, and \cite{griffin86}.
Also plotted in the figures are two representative model temperature spectra,
together with the series of molecular lines from which they are derived.  
The Jupiter models are from \cite{griffin86}, and assume
clear-sky (dashed) or NH$_3$ cloud cover with particle size 100~$\mu$m and a 
particle scale height to gas scale height ratio of 0.15 (dotted).
The Saturn models are from \cite{hildebrand85}, and assume 
an NH$_3$ mixing ratio of $2\times 10^{-4}$ in the deep atmosphere, and
PH$_3$ mixing ratios equal to $1.5\times 10^{-6}$ (dashed) or 
$1.0\times 10^{-5}$ (dotted).

Note that our results at 16.4~\icm\ for both Jupiter and Saturn are 
significantly lower than previous measurements that cover this band.
While we do not completely understand the cause of this, we offer several
observations:
The bandwidth of our 16.4~\icm\ filter is significantly narrower than 
the previous measurements (due to \cite{hildebrand85}).  This band is 
nearly coincident with the first expected strong dip in the spectra
of the giant planets, near the $\sim$ 19~\icm\ NH$_3$ and PH$_3$ resonances.
Additionally, the measurements reported here had essentially 100\%
atmospheric transmission.
The earlier measurements were made from the ground on Mauna Kea, and 
were corrected to a fixed value of the line of sight water vapor before
taking ratios of unknown to calibration signals.  

\section{Conclusions}

We have reported a new set of whole-disk brightness ratios for Mars, 
Jupiter, and Saturn which, when combined with a thermal model of Mars,
give new calibration values for Jupiter and Saturn.  The overall uncertainties
on the brightness temperatures are still dominated by model uncertainty
in the mm/sub-mm emission of Mars, but the ratio measurements presented
here should remain useful well into the future, permitting a straightforward
revision of the brightness temperatures of the giant planets with improvements
in the Martian model.  
As was mentioned in \cite{hildebrand85}, the 
effective mm/sub-mm emission radii of the gas giants are uncertain by as much
as 1\%, but while this uncertainty is important for understanding the
precise brightness temperature of the planet, it does not affect the total
flux density, and thus is not an issue for use in large-beam calibration
studies.

\acknowledgments
We would like to thank E. Weisstein and D. Muhleman for several
useful discussions on the thermal model for Mars.
We also thank the Free Software Foundation for the GNU software tools, 
and J.~W. Eaton for Octave, both of which were extensively used in this work.  
This research was supported by the National Aeronautics and Space 
Administration through the Office of Space Science.

\clearpage

\end{document}